\documentclass[twocolumn,amsmath,amssymb,floatfix, groupedaddress, prx]{revtex4-2}
\usepackage{leading}
\usepackage[latin3]{inputenc}
\usepackage[makeroom]{cancel}
\usepackage{graphicx}
\usepackage{amsmath}
\usepackage{amsfonts}
\usepackage{amssymb}

\usepackage{color}
\usepackage[left=2cm,right=2cm,top=2cm,bottom=2cm]{geometry}
\usepackage{graphicx} 
\usepackage{dcolumn}
\usepackage{bm}
\usepackage{simplewick}
\usepackage{array}
\usepackage{appendix}
\usepackage{cancel}
\usepackage[normalem]{ulem} 
\RequirePackage[
   hyperindex,colorlinks,bookmarksnumbered,
   plainpages=true,pdfstartview=FitH]{hyperref}
\hypersetup{linkcolor=blue,urlcolor=blue,citecolor=blue}
\usepackage{hyperref}
\usepackage{mathrsfs}

\definecolor{darkblue}{rgb}{0, 0, 0.8}

\newcommand{\lr}[1]{\left( #1 \right)}

\newcommand{\ket}[1]{\left| #1 \right>} % for Dirac bras
\newcommand{\bra}[1]{\left< #1 \right|} % for Dirac kets
\newcommand{\braket}[2]{\left< #1 \vphantom{#2} \right|\left. #2 \vphantom{#1} \right>} 
\begin{document}
\title{Physics of the Majorana-superconducting qubit hybrids}
\author{D. B. Karki, K. A. Matveev, and Ivar Martin}
\affiliation{Material Science Division, Argonne National Laboratory, Argonne, IL 08540, USA}

\begin{abstract}
Manipulation of decoupled Majorana zero modes (MZMs) could enable topologically-protected quantum computing. However, the practical realization of a large number of perfectly decoupled MZMs needed to perform nontrivial quantum computation has proven to be  challenging so far.
Fortunately, even a small number of imperfect MZMs can be used to qualitatively extend the behavior of standard superconducting qubits, allowing for new approaches for noise suppression, qubit manipulation and read-out. Such hybrid devices take advantage of interplay of Cooper pair tunneling, coherent single electron tunneling, and Majorana hybridization. 
Here we provide a qualitative understanding of this system, give analytical results for its ground state energy spanning full parameter range, and describe potential sensing applications enabled by the interplay between Majorana and Cooper pair tunneling. 
\end{abstract}

\maketitle
\section{Introduction}
Majorana zero modes (MZMs) are the key element of proposed topological quantum computers~\cite{Majorana1937, Nayak_2008}. 
In the presence of MZMs, the ground state acquires topological degeneracy and manipulation of MZMs allows it to effect non-trivial transformations in the multidimensional ground-state manifold~\cite{Nayak_2008}. While MZMs are naturally expected to appear at the edges or inside vortex cores in the topological superconductors~\cite{ReadGreen_2000, Avanov_2001, Kitaev_2001, Kitaev_2003}, the scarcity of such bulk materials has led to alternative theoretical proposals for realization and manipulation of MZMs based on superconducting proximity effect~\cite{FuKane_2008, FuKane_2009, Sau_Sarma_2010, Alicea_2010, Lutchyn_Sarma_2010, Oreg_vOppen_2010}. Despite significant experimental advances, which report certain features consistent with MZMs~\cite{Mourik_2012,Heiblum_2012,Finck_2013, Marcus_2016, Marcus_2016a, Kouwenhoven_2018, Yacobi_2019,Nayak_2023}, convincingly realizing  MZMs remains a challenge~\cite{Yacoby_2023}. 

For topological quantum computation, MZMs have to be well spatially isolated from each other. However, in practice MZMs are expected to be exponentially localized on the scale of the superconducting coherence length~\cite{splitting1, splitting2}, which can be quite long, particularly when topological superconductivity is induced via proximity effect. Achieving decoupled MZMs thus requires large clean systems, which is very difficult to achieve in practice. Implementing MZMs in smaller systems appears feasible; however, finite Majorana hybridization cannot be ignored in that case.

When MZMs are not decoupled, they cannot be easily used for topological quantum computing, even though certain purification procedures can be applied to dynamically decouple them~\cite{AM1}. Fortunately, introducing even imperfect (coupled) MZMs into other quantum devices can lead to new interesting phenomena and functionalities. 
Indeed, hybrids of MZMs with the more standard superconducting quantum technology have been gaining attention recently~\cite{DasSarma_2015,Plugge_2017, Karzig_2017, vOppen_2020}. Various ideas have been put forwarded differing mainly in the type of base qubit, ranging from the use of flux qubit~\cite{fluxq, Hassler_2010}, charge qubit~\cite{Hassler_2011}, fluxonium~\cite{Demler_2013} and transmon~\cite{Ginossar_2014, Ginossar_2015}. 
\begin{figure}[b]
\begin{center}
\includegraphics[scale=0.5]{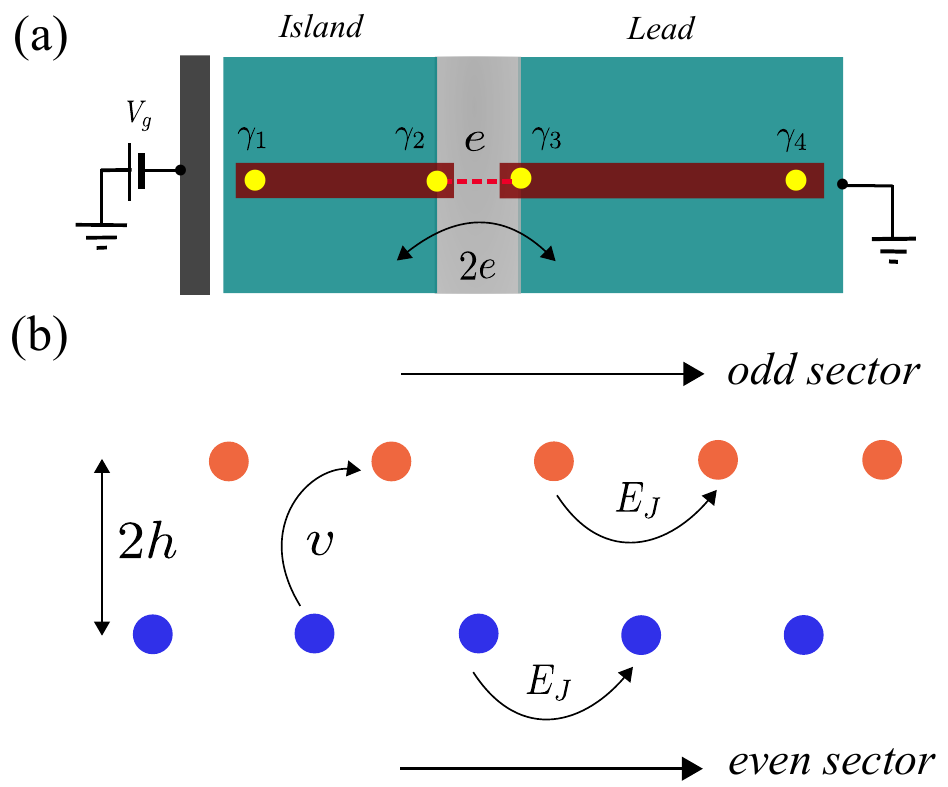}
\caption{(a) The schematic representation of MZMs-SC qubit hybrid setup under consideration (see the text). (b) The diagrammatic illustration of the model Hamiltonian given in Eq.~\eqref{seteq0}. Orange and blue are different number states in odd and even parity sectors respectively, separated by an energy $2h$, with $h$ being Majorana overlap. Josephson (Cooper pair) tunneling $E_J$ preserves the parity sector, connecting states  $\ket n$ and $\ket{n+2}$; Majorana tunneling $v$ connects states $\ket n$ and $\ket{n+1}$ in the two sectors. }\label{setup1}
\end{center}
\end{figure}

When combined with superconducting (SC) qubits, MZMs significantly extend their functionality by modifying the current-phase relationship and introducing a new degree of freedom, the fermion parity. The extra tunability of this hybrid setup allows for additional means of suppression of environmental noises and new qubit manipulation and read out protocols~\cite{Keselman_2019, Prada_Aguado_2020a, Prada_Aguado_2020b, Luca_2022}.
However, due to the complexity of the interplay between coherent single electron tunneling, pair tunneling and the  MZMs hybridization in the MZMs-SC qubit hybrid, this problem has been tackled mostly numerically in the previous works~\cite{Keselman_2019, Prada_Aguado_2020a, Prada_Aguado_2020b}.

In this work, we focus on the qualitative aspects of the hybrid MZMs-SC qubit setup. We provide a simple physical way to understand this system based purely on the coherent charge dynamics, which avoids the need of more subtle considerations of the quantum wave function boundary conditions used in the previous works. We also develop an analytical method for obtaining the ground state energy of the system. We find that this hybrid device has a rich variety of operational regimes. We discuss their characteristic signatures and the crossovers between them. We highlight the practical applications of the proposed hybrid device and show how it could contribute to the unambiguous detection of Majorana zero modes.

The organization of the paper is as follows. The section~\ref{setupmodel} introduces general features of MZMs-SC qubit hybrid along with its Hamiltonian formulation. Different parity sectors of this setup are discussed in Sec~\ref{dpsbc}. Here we provide  a simple physical way to understand periodic and anti-periodic boundary conditions invoked in Ref.~\cite{Ginossar_2014} to correctly solve the Hamiltonian of MZMs-SC qubit setup. We describe  the qualitative features of the exact spectrum in Sec.~\ref{qfeatures}. The section~\ref{sec:an} is devoted to the presentation of analytical calculations for the lowest energy state of the hybrid setup. In Sec.~\ref{blockade}, we discuss the emergent Josephson physics in the proposed hybrid device and suggest some practical applications. Finally, we conclude in Sec.~\ref{dissoutlook}. Mathematical details of our calculations are deferred to the Appendices.

\section{Setup and model}\label{setupmodel}
We consider a ``floating" superconducting island with charging energy $E_C$,  gated capacitively by voltage $V_g$ as shown in Fig.~\ref{setup1}(a). The island is tunnel coupled to the superconducting lead via the Josephson coupling $E_J$, which allows Cooper pairs to coherently tunnel between the island and the lead. In addition, there are two nanowires, each hosting a pair of MZMs $\gamma_{1, 2}$ and  $\gamma_{3, 4}$, attached to the island and the lead, respectively. The parities of the fermion number of wires are given by $i\gamma_1\gamma_2$ and  $i\gamma_3\gamma_4$. Tunnel coupling $v$ between between $\gamma_2$ and $\gamma_3$, allows a coherent transfer of individual electrons between the island and the lead. While it flips the individual wire parities, it preserves the global fermion parity of the system, $\gamma_1\gamma_2\gamma_3\gamma_4$, which can be either odd or even, forming two disconnected sectors. In each sector, we can define Pauli operators that span the remaining two-dimensional Hilbert space, $\sigma_z = i\gamma_1\gamma_2$, $\sigma_x = i\gamma_2\gamma_3$, and $\sigma_y = i\gamma_3\gamma_1$.

Using this representation, the Hamiltonian is
\begin{eqnarray}
H&=&E_C (n - n_g)^2  + h \sigma_z \nonumber\\
&&- \frac v 2\sigma_x \lr{\ket{n+1}\bra{n} + \ket{n}\bra{n+1}}  \nonumber\\
&&- \frac {E_J}{2} \lr{ \ket{n+2}\bra{n} + \ket{n}\bra{n+2}}.\label{seteq0} 
\end{eqnarray}
Here $n$ counts the number of electrons on the island and $n_g$ is the offset charge, which is proportional to the gate voltage $V_g$. The term proportional to $h$ takes into account the hybridization between $\gamma_1$ and $\gamma_2$ on the wire attached to the island (due to the conservation of the total parity, the hybridization between $\gamma_3$ and $\gamma_4$ has the same form).  
The Hamiltonian~\eqref{seteq0} can be equivalently expressed in eigenbasis of the phase operator, conjugate to the particle number,   $n=-i\partial_\varphi$, such that $\braket{n}{\varphi} = e^{in\varphi }$. In this basis the tunneling is diagonal, however the charging energy becomes a differential operator,
\begin{equation}
H=E_C \lr{{-i\partial_\varphi} - n_g}^2 +h\sigma_z-v \sigma_x\cos\varphi - E_J \cos2 \varphi.\label{seteq1}
\end{equation}

Compared to the standard Hamiltonian for a Cooper pair box, in addition to the pair tunneling, now {\em coherent} single electron tunneling is also allowed. Note that in conventional superconductors single-electron tunneling is always incoherent since it corresponds to creation or destruction of an unpaired quasiparticle in the continuous spectrum. The presence of MZMs enables coherent single electron tunneling. The spectral isolation of MZMs implies that the state of the system can be completely described by the number of electrons on the island and the internal state of the two level system associated with four MZMs. This is a qualitative difference between the standard Josephson qubit devices, and the Majorana-enriched devices, which leads to new opportunities in the spectral engineering and quantum state control.

\section{Parity sectors and boundary conditions}\label{dpsbc}
From the Hamiltonian~\eqref{seteq0} it is clear that there is a relationship between $\sigma_z$ and the parity of island charge $n$, with $\sigma_z (-1)^n$ being conserved by the Hamiltonian. 
For instance, if we  assume that we are in the sector $\sigma_z (-1)^n = 1$, then, in the absence of tunneling, the spectrum is given by $E_n = E_C (n - n_g)^2  + (-1)^n h $ and the wave functions for the two parities of $n$ by $ \ket{2k}\ket{\uparrow}$ and $\ket{2k+1}\ket{\downarrow}$.
%with the corresponding wave functions, different for the two parities of $n$,  $\Psi_{2k} = \ket{2k}\ket{\uparrow}$ and $\Psi_{2k+1} = %\ket{2k+1}\ket{\downarrow}$. 
%The Josephson term does not mix different parities of $n$; however, single electron tunneling, enabled by Majoranas, does. 

The parity constraints in representation (\ref{seteq1}) are more  obscured. To reveal them, consider again the limit of zero MZM and Josephson tunneling, i.e., $v=E_J=0$. The eigenstates are expected to be $e^{i n \varphi} \ket{\sigma_z}$, with the energies $E_n = E_C (n - n_g)^2  + h\sigma_z$. However, as we saw above, the value of $\sigma_z$ is pinned to the parity of $n$. Therefore, the physical wave functions have the form  $\lr{\begin{array}{c} e^{i 2k \varphi}\\0\end{array}}$ and $\lr{\begin{array}{c} 0\\e^{i (2k+1) \varphi}\end{array}}$. Note that both functions have period $2\pi$; however, the former is also periodic on interval of length $\pi$ and the latter is antiperiodic on that interval. 
When both the Majorana and Josephson tunnelings are turned on, the wave function becomes a superposition of all allowed charge states and thus
acquires the form $\Psi_\varphi = \lr{\begin{array}{c} g_\varphi\\f_\varphi \end{array}}$, with general $g_{\varphi +\pi} = g_{\varphi}$ and $f_{\varphi +\pi} = - f_\varphi$.  This observation, also made previously but using a different reasoning~\cite{Ginossar_2014, Keselman_2019, Prada_Aguado_2020a, Prada_Aguado_2020b} is important for avoiding non-physical states -- imposing only the periodicity on the full interval $2\pi$ retains both $\sigma_z (-1)^n = \pm 1$ sectors, which is not physical.

\subsection*{Hamiltonian in a fixed $\sigma_z (-1)^n = -1$ sector}\label{sec:phH}
In this section we go a step further, and show that given the structure of the Hamiltonian (\ref{seteq0}) which pins the parity of charge and the spin associated with the MZMs, it is possible to eliminate the spin degree of freedom and associated redundancy completely. Without loss of generality, in the sector $\sigma_z (-1)^n = -1$, Hamiltonian becomes
\begin{eqnarray}
H&=&E_C (n - n_g)^2  - h(-1)^n \nonumber\\
&&- \frac v 2 \lr{\ket{n+1}\bra{n} + \ket{n}\bra{n+1}}  \nonumber\\
&&- \frac {E_J}{2} \lr{ \ket{n+2}\bra{n} + \ket{n}\bra{n+2}}.\label{seteq0wos} 
\end{eqnarray}
The staggered potential evokes an analogy with a charge density wave ordering in electronic systems. The charging energy $E_C$ breaks the translational invariance in $n$ space, superimposing a parabolic confining potential. 

 The wave function $\ket \psi = \sum_n \psi_n  \ket{n}$ satisfies the standard Shr{\"o}dinger equation, $E\psi  = H \psi$,
\begin{align}
E \psi_n &=
\left[E_C\left(n-n_g\right)^2-h(-1)^n \right]\psi_n\nonumber\\
&\;\;\;\;\;-\frac{v}{2}\left(\psi_{n+1}+\psi_{n-1}\right)\nonumber\\
&\;\;\;\;\;-\frac{E_J}{2}\left(\psi_{n+2}+\psi_{n-2}\right).\label{eq:SEn}
\end{align}

Instead of the particle number  basis, we can specify the wave function in the conjugate phase space.  We chose the transformation between the two bases as  
\begin{equation}
\psi_n=\frac{1}{\sqrt{2\pi}}\int^\pi_{-\pi}d\varphi\psi(\varphi)e^{-i(n-n_g)\varphi}.\label{wkbeq20}
\end{equation}
It is convenient to explicitly include the $n_g$ shift in the exponent, which corresponds to imposing the twisted boundary conditions, $\psi(\varphi+2\pi)=e^{-2\pi i n_g}\psi(\varphi)$; this eliminates $n_g$ from the {Hamiltonian.  In terms of the function $\psi(\phi)$, the Schr\"odinger equation~\eqref{eq:SEn} takes the form}
\begin{align}
E \psi(\varphi) =&-\left(E_C\partial^2_\varphi+v\cos\varphi+E_J\cos 2 \varphi\right)
\psi(\varphi)\nonumber\\
&-h\psi\left(\varphi-\pi{\rm sgn}\varphi\right)e^{-i\pi n_g{\rm sgn}\varphi},\label{eq:SEphi}
\end{align}
where we assume $-\pi\leq\varphi\leq\pi$.
It is now apparent that the role of the staggered potential $h$ is to introduce coupling between the Fourier components $\phi$ and $\phi+\pi$ of $\psi(\varphi)$, in analogy with the charge density waves. 

For further simplification, we define a two-component function $\Psi(\varphi)$ as
\begin{align}
\Psi(\varphi)=\begin{pmatrix}\Psi_{\uparrow}(\varphi)\\\Psi_{\downarrow}(\varphi)\end{pmatrix},\label{twocompf}
\end{align}
where $\Psi_{\uparrow,\downarrow}$ are defined in the half interval $0\leq \varphi\leq\pi$, and are related to $\psi$ by
\begin{align}
\Psi_{\uparrow}(\varphi)&=\psi(\varphi+\pi)e^{i\pi n_g},\nonumber\\
\Psi_{\downarrow}(\varphi)&=\psi(\varphi).\label{wkbeq30b}
\end{align}
The boundary conditions are now imposed on the $[0,\pi]$ interval, mixing  $\Psi_{\uparrow,\downarrow}$ such that
\begin{align}\label{eq:bc}
\Psi_\uparrow(\varphi+\pi)&=\Psi_\downarrow(\varphi) e^{-i\pi n_g},\nonumber\\
\Psi_\downarrow(\varphi+\pi)&=\Psi_\uparrow(\varphi) e^{-i\pi n_g}.
\end{align}
In this representation, Eq.~\eqref{eq:SEphi} takes the convenient form
\begin{align}
&-\left(E_C\partial^2_\varphi+E_J\cos 2 \varphi-v\cos\varphi\right)\Psi_\uparrow - h\Psi_\downarrow = E \Psi_\uparrow\nonumber\\
& -\left(E_C\partial^2_\varphi+E_J\cos 2 \varphi+v\cos\varphi\right)\Psi_\downarrow- h\Psi_\uparrow = E \Psi_\downarrow \nonumber.
\end{align}
More compactly, these equations can be expressed in the form
\begin{align}
\mathbb{M}\Psi=E\Psi,\label{wkbeq31}
\end{align}
where the matrix $\mathbb{M}$ is given by
\begin{equation}
\mathbb{M}=-\left(E_C\partial^2_\varphi+E_J\cos 2\varphi\right)+v\cos\varphi\sigma_z-h\sigma_x.\label{wkbeq31a}
\end{equation}
In the following, we will seek the solution of the Schr{\"o}dinger equation~\eqref{wkbeq31} with the boundary conditions~(\ref{eq:bc}). This representation eliminates the redundancy associated with the disconnected sectors $\sigma_z (-1)^n = \pm 1$, which are both  present in Eq.~\eqref{seteq1}. In Appendix~\ref{app:33} we provide a correspondence between this approach and the one followed in~\cite{Ginossar_2014} and related works.

\section{Qualitative features of the exact spectrum}\label{qfeatures}
\begin{figure}[t]
    \centering
    \includegraphics[scale=0.38]{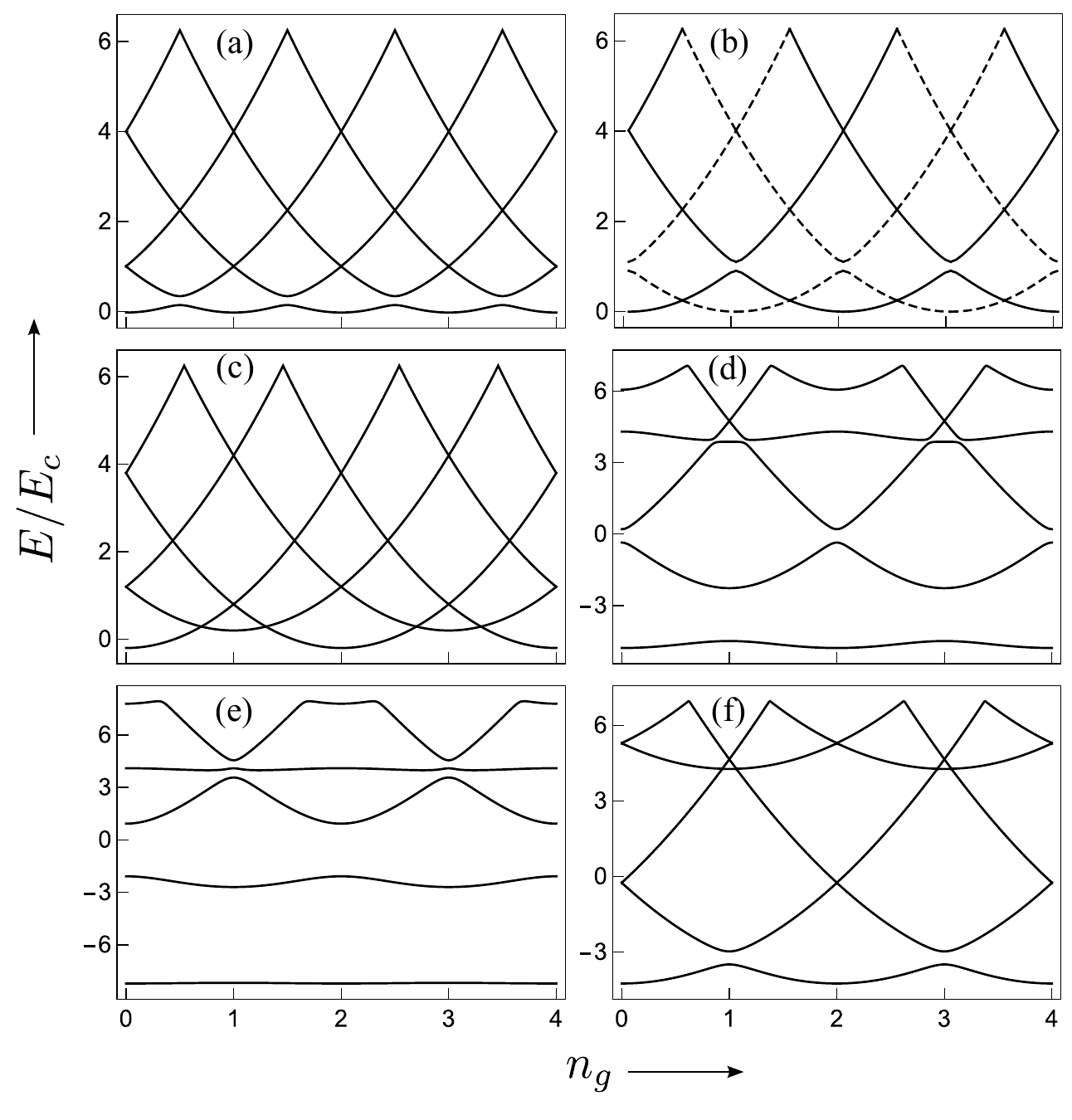}
   \caption{The first five energy levels of the spectrum as a function of offset charge obtained from the numerical solution of Eq.~\eqref{seteq0}. The plots (a)--(f) correspond to $(h/E_C,\; v/E_C,\; E_J/E_C)$=(0, 0.2, 0), (0, 0, 0.2), (0.2, 0, 0), (4, 2, 2), (4, 5, 5), (4, 2, 0.25) respectively.}\label{fig-spectra}
\end{figure}

Before presenting  the analytical results, in this brief section we discuss some representative features of the MZMs-SC qubit system. For our numerical calculations we use Eq.~\eqref{eq:SEn}.  Several examples of the spectra are shown in Fig.~\ref{fig-spectra}. All energies and parameters are normalized by the charging energy $E_C$. 
In panel (a) the only other nonzero parameter (apart from $E_C$) is the Majorana tunneling $v$. The spectrum as a function of $n_g$ has the overall characteristic form of Coulomb parabolas $(n - n_g)^2$, $n$ being all integers, with the avoided crossings that decay exponentially as $\sim (v/E_C)^{|n_1 -n_2|}$.  When $E_J \ne 0$ and $v = 0$, as in the panel (b), the avoided crossings appear only between parabolas that correspond to the even differences of $n$. The two sets of bands that correspond to different parities of $n$ are present simultaneously (solid and dashed lines, respectively), without hybridization.  That is, unless the island parity is allowed to be flipped by finite $v$, the dispersion has period 2 in $n_g$. 

Panel (c) illustrates the effect of finite MZMs hybridization within the island, $h$, assuming that tunneling between the island and the lead is turned off. These are simple Coulomb parabolas, but now the odd and even parabolas are offset in energy. 

Panels (d)--(f) show what happens when the MZMs hybridization $h$ is large compared to the charging energy. The period 2 in $n_g$ becomes apparent. Another notable feature is that for sufficiently strong $E_J, |v| \gtrsim 1$, there are two weakly dispersing bands, one associated with the ground states in each parity sector (approximately separated by $2h$ in energy). This is in contrast to the standard transmons, where there is only one such band, at the lowest energy. Within a given parity sector, the system approximately behaves as a transmon, but with an effective Josephson coupling {$\tilde E_J = E_J - \sigma_z v^2/4h$},  and the lowest band width scaling as $e^{-\sqrt{32\tilde E_J/E_C}}$. Interestingly, in the upper band, $\tilde E_J$ can vanish when the MZM-mediated tunneling exactly offsets the standard Josephson pair tunneling. This corresponds to a unique situation when the fermion parity on one side of the junction can drastically affect the Josephson critical current, which potentially can be used as a smoking gun for the presence of MZMs, or for charge/photo sensing applications. Further discussion of this effect is presented in Sec.~\ref{blockade}.

\section{Analytical calculations for the lowest energy state}\label{sec:an}

In this section we turn our attention to the properties of the lowest-energy band, $E_0(n_g)$. To isolate the unique features associated with MZMs, we will set $E_J = 0$, which eliminates the standard Cooper pair tunneling.  For $h = 0$ in this case we should obtain $E_0(n_g + 1) = E_0(n_g)$. This charge-translation symmetry is broken, however,  by any finite $h$, leaving only the $E_0(n_g + 2) = E_0(n_g)$ symmetry intact. This is the same symmetry as in the conventional superconductor transmons and Cooper-pair boxes, even though we are not allowing standard pair tunneling; instead, it is a result  of the Majorana hybridization $within$ the island, which makes the energies of the even and odd charge states different. 

Our focus will be on the case $v/E_C \gg 1$, which corresponds to small charging energy. In this case, the effects of gate charge fluctuations are strongly screened, making this an attractive regime for MZMs-SC qubit. The dominant energy scale is associated with the Majorana tunneling, which tends to pin the phase near $\varphi = 0$, see Eq.~(\ref{eq:SEphi}). The role of the nonzero charging energy $E_C$ is to provide fluctuations around this value, and to make the energy sensitive to the twist in the boundary conditions, $\psi(\varphi+2\pi)=e^{-2\pi i n_g}\psi(\varphi)$. This creates a finite bandwidth for $E_0(n_g)$. In the language of tight-binding, the charging energy gives a finite mass to the particles, allowing them to tunnel between the potential minima, located at $\phi = 2\pi n$, with $n$ integer. The gate charge $n_g$ plays the role of the quasimomentum associated with this lattice, as can be seen from the form of the boundary conditions. 

For $h = 0$, there is no formal difference between the Majorana-based model, and the standard Cooper-pair transmon, apart from rescaling all charges by a factor of two (single electron charge vs. Cooper pair charge).
At finite $h$, the situation changes dramatically. There is no natural analog of $h$ in the case of Cooper pairs, it is a very special feature of a system with Majoranas that the energy can depend not on charge, but solely on the $parity$ of charge. This leads to the appearance of several qualitatively distinct regimes, separated by crossovers.

\subsection{Transmon regime: zero Majorana overlap}
\begin{figure}[t]
\begin{center}
\includegraphics[scale=0.5]{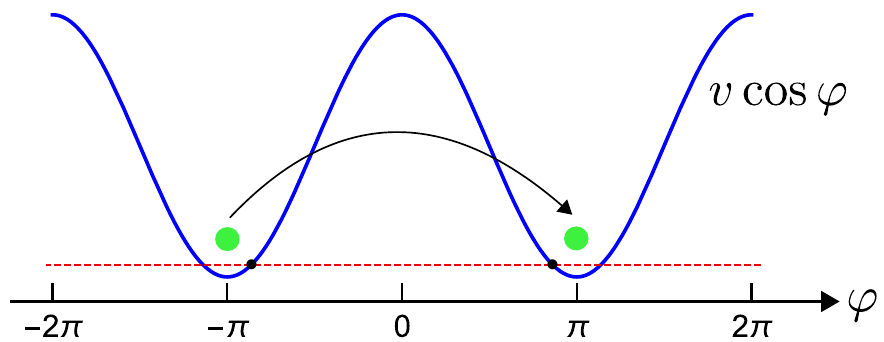}
\caption{Illustration of instanton tunneling in the cosine potential. The frequency $\Omega$ of small amplitude oscillation around the minimum $\varphi=\pm \pi$ is defined by $m\Omega^2=v$. In the limit of $v\gg E_C$, the right and left turning point of the potential $v\cos\varphi$ are $\varphi=\pm\pi\mp z$, with $z^2\simeq \hbar/(m\Omega)$. The zero point energy is represented by the dashed line.}\label{instfig}
\end{center}
\end{figure}

In the ideal case of zero Majorana overlap $h$, the Schr\"odinger equation becomes the standard Mathieu equation with cosine potential,
\begin{align}
E \psi(\varphi) &=-\left(E_C\partial^2_\varphi+v\cos\varphi\right)
\psi(\varphi)\label{eq:SEphi2}.
\end{align}

Transparent analytical results for the ground state energy can be obtained in the strong barrier limit  $v/E_C\gg 1$, by using the semiclassical Wentzel-Kramers-Brillouin (WKB) method. The WKB regime can also be understood in terms of  the instanton tunneling~\cite{Coleman_1997} events between the neighboring minima of the cosine potential as illustrated in Fig.~\ref{instfig}. The phase in the twisted boundary conditions then acquires the meaning of the lattice quasimomentum. The instanton tunneling produces the well-known energy splitting $\Delta^{(0)}$, which is given by~\cite{Koch_2007}
\begin{align}
\frac{\Delta^{(0)}}{E_C} &=\frac{2^{\frac{13}{4}}}{\sqrt{\pi}}\left(\frac{v}{E_C}\right)^{\frac{3}{4}}\exp\left(-4\sqrt2\sqrt{\frac{v}{E_C}}\right).\label{wkbeq4}
\end{align}

This tunneling matrix element determines the width of the lowest energy band. Explicitly, reintroducing the ``quasimomentum" $n_g$, ground state energy takes the form
\begin{equation}
    E_0 = -\Delta^{(0)} \cos(2\pi n_g) + {\rm const}.
\end{equation}
The energy has period $1$ in $n_g$, thanks to the single-electron tunneling enabled by the presence of Majorana fermion states. This should be contrasted with the case of the standard superconducting junctions where only Cooper pairs can tunnel coherently and hence the energy has period 2 as a function of $n_g$.

\subsection{Perturbative regime: small Majorana overlap}
In this section, we account for the small Majorana overlap $h$ perturbatively. To this end, we use  Eq.~\eqref{wkbeq31} and treat the second term as a perturbation to get the linear in $h$ correction to the ground state energy 
\begin{align}
\delta E_0=h\left.\frac{\partial\left<\mathbb{M}\right>}{\partial h}\right|_{h\to 0}.\label{wkbeq5}
\end{align}
Using equations~\eqref{wkbeq5}, ~\eqref{wkbeq31a} and~\eqref{wkbeq30b}, it is straightforward to show that
\begin{align}
\delta E_0=-h\int^{\pi}_{-\pi}d\varphi\psi_0^*(\varphi)\psi_0(\varphi-\pi{\rm sgn}\varphi) e^{-i\pi n_g {\rm sgn}\varphi}
.\label{wkbeq5aa}
\end{align}
Here the subscript in $\psi_0$ represents the limiting case of $h=0$. In the limit of $v/E_C\gg1$, the wave function $\psi_0$ can be calculated using semiclassical approximation. As detailed in the appendix~\ref{app1}, the leading order contribution to $\delta E_0$ takes the form
\begin{align}
\delta E_0=-\Delta^{(1)}\cos \pi n_g,\label{wkbeq5a}
\end{align}
where 
\begin{align}
\frac{\Delta^{(1)}}{E_C}= \left(\frac{ h}{E_C}\right)2^{\frac{11}{4}}(\sqrt2-1)\exp\left[{-}4(\sqrt2-1)\sqrt{\frac{v}{E_C}}\right].\label{wkbeq12}
\end{align}

Using equations~\eqref{wkbeq4} and~\eqref{wkbeq5a}, we get the ground state energy 
\begin{equation}\label{energygnd}
E_0=-\Delta^{(0)}\cos 2\pi n_g-\Delta^{(1)}\cos \pi n_g+{\rm const}.
\end{equation}
Notably, while the first term has period 1 in $n_g$, the second contribution due to $h$ has period 2. From equations~\eqref{wkbeq4} and~\eqref{wkbeq12}, it is readily seen that for exponentially small Majorana overlap
\begin{equation}
h/E_C\simeq(v/E_C)^{3/4}\exp\left(-4\sqrt{v/E_C}\right),\label{hc}
\end{equation}
the tunnel splittings $\Delta^{(0)}$ and $\Delta^{(1)}$ are of the same order of magnitude. Therefore, already for exponentially small $h$, the energy dependence on the gate charge has the same period 2 as in the case of conventional superconductors. This can make the determination of the presence of MZMs in the experimental devices of this kind challenging. 

We note that the splitting relation~\eqref{wkbeq12} has been derived using the first order perturbation theory. It is not yet obvious what is the regime of its applicability. This issue is addressed below in Sec.~\ref{crossoverreg}.

\subsection{WKB regime: intermediate and large Majorana overlap}\label{wkbregm}
\begin{figure}[h]
\begin{center}
\includegraphics[scale=0.5]{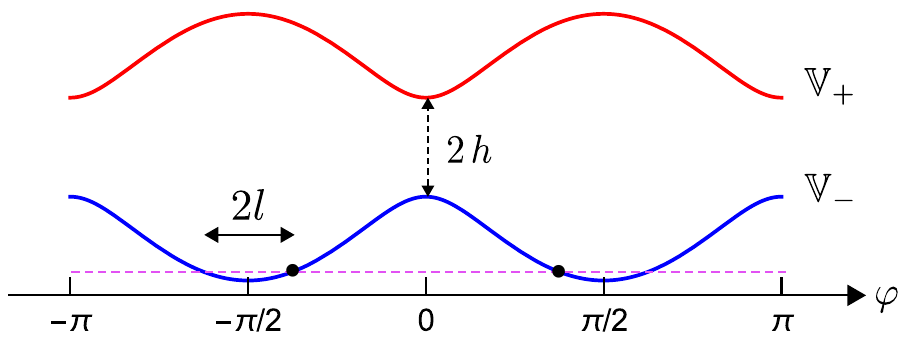}
\caption{Illustration of WKB regime where the Majorana overlap $h$ is sufficiently large such that the upper potential branch can be disregarded. The potential $\mathbb{V}_-$ has right and left classical turning points at $\varphi=\pm\pi/2\mp l$. The zero point energy is represented by the dashed line (see text for the details).}\label{wkbfig}
\end{center}
\end{figure}

The Hamiltonian~\eqref{wkbeq31a} can be interpreted as a spin-1/2 interacting with a non-linear oscillator. When the oscillator dynamics is slow compared to the ``Zeeman" field acting on the spin, the latter simply follows the field. This allows to integrate out the spin, assuming that it is either aligned or antialigned with the Zeeman field. The corresponding equation of motion for the oscillator is
\begin{equation}\label{sum1}
\frac{\partial^2\Psi(\varphi)}{\partial\varphi^2}+\frac{1}{E_C}\left[E-\mathbb{V}(\varphi)\right]\Psi(\varphi)=0.
\end{equation}

The effective potential $\mathbb{V}$ obtained from Eq.~\eqref{wkbeq31a} consists of two branches
\begin{eqnarray}
  \mathbb{V}_\pm&=&v\left(\sqrt{\alpha^2+1}\pm\sqrt{\alpha^2+(\sin\varphi)^2}\right),\;\\
\alpha&=&h/v,\label{w1}  
\end{eqnarray}
where we set $E_J = 0$ as before, shifted the phase variable by $\pi/2$ and added the constant $\sqrt{h^2+v^2}$ for convenience, see Fig. \ref{wkbfig}. The separation between the potential branches is the above-mentioned phase-dependent Zeeman field. It reaches minimum value of $2h$ at $\varphi = 0$. We thus expect the adiabatic approximation to hold for sufficiently large $h$, which is to be determined in the following.

In the adiabatic limit, the upper potential branch $\mathbb{V}_+$ can be ignored for the calculation of ground state splitting. In the strong barrier limit, the usual WKB approach can be exploited for the evaluation of the tunnel splitting $\Delta^{(2)}$, which is given by (see appendix~\ref{wregime} for details)
\begin{align}
\frac{\Delta^{(2)} }{E_C}&=2^{\frac{1}{4}}\sqrt\pi\left(\frac{v}{E_C}\right)^{\frac{3}{4}}\left(\frac{1}{1+\alpha^2}\right)^{\frac{3}{8}}e^{\mathcal{S}_0}\;e^{\mathcal{S}_1}.\label{w3}
\end{align}
Here $\mathcal{S}_0$ and $\mathcal{S}_1$ are expressed in terms of standard Hypergeometric functions as
\begin{align}
\mathcal{S}_0&={-}2\sqrt{\frac{v}{E_C}}\!\left(\frac{1}{4(1{+}\alpha^2)}\right)^{1/4}\!\!\, _3F_2\left(\frac{1}{4},\frac{3}{4},1;\frac{3}{2},\frac{3}{2};\frac{1}{\alpha ^2{+}1}\right),\nonumber\\
\mathcal{S}_1&=\log \left(\!\frac{4}{\pi }\!\right){-}\frac{1}{8(1{+}\alpha^2)}\, _4F_3\!\left(\frac{3}{4},1,1,\frac{5}{4};\frac{3}{2},\frac{3}{2},2;\frac{1}{\alpha ^2{+}1}\right).\label{w4}
\end{align}

Equation~\eqref{w3} in the limit of large $\alpha$ yields the splitting of the form
\begin{equation}
\frac{\Delta^{(2)}_a}{E_C}=\frac{2^{\frac{9}{4}}}{\sqrt{\pi}}\left(\frac{v^2}{hE_C}\right)^{\frac{3}{4}}\exp\left(-\sqrt{\frac{2v^2}{hE_C}}\right).\label{w9}
\end{equation}
This result is closely related to the band width of the lowest state in the conventional transmon, if we replace the Josephson energy by $v^2/4h$. This is indeed expected since for large $\alpha$, transferring an electron from the island costs energy $h$, which has to be offset either by returning electron to the island or by tunneling a second electron from the island to the lead. This is in direct analogy to the Josephson coupling which originates from the processes where Cooper pairs are transiently broken in the process of tunneling; however, there the  energy of the virtual state is given by the superconducting gap, instead of $h$.

The small $\alpha$ limit of Eq.~\eqref{w3} is given by
\begin{align}
\frac{\Delta^{(2)}_b }{E_C} &=\frac{2^{\frac{13}{4}}(\sqrt2-1)}{\sqrt\pi}\left(\frac{v}{E_C}\right)^{\frac{\eta}{2}+\frac{3}{4}}\left[4\left(\sqrt2-1\right)^2\right]^{2\eta}\nonumber\\
&\;\;\;\;\;\;\;\;\;\;\times e^\eta \eta^{-\eta}\;\exp\left(-4(\sqrt2-1)\sqrt{\frac{v}{E_C}}\right),\label{w6a}
\end{align}
where we introduced additional control parameter $\eta$, which is defined by
\begin{equation}
\eta=\frac{h^2}{4\sqrt{E_C v^3}}.\label{w7}
\end{equation}
It is readily seen that the small $\alpha$ limit of the splitting given by Eq.~\eqref{w6a} is not the same as the linear in $h$ result obtained earlier in Eq.~\eqref{wkbeq12}. This is expected  since for the  WKB calculation to remain valid, $h$ needs to be sufficiently large such that the upper potential branch can be neglected. Therefore, there should exist a crossover regime that connects the WKB and perturbative regimes. In the following section, we study this crossover regime and show that the parameter $\eta$ defined in Eq.~\eqref{w7} indeed serves as the crossover parameter.

\subsection{Crossover between the perturbative and WKB regimes}\label{crossoverreg}
\begin{figure}[b]
\begin{center}
\includegraphics[scale=0.5]{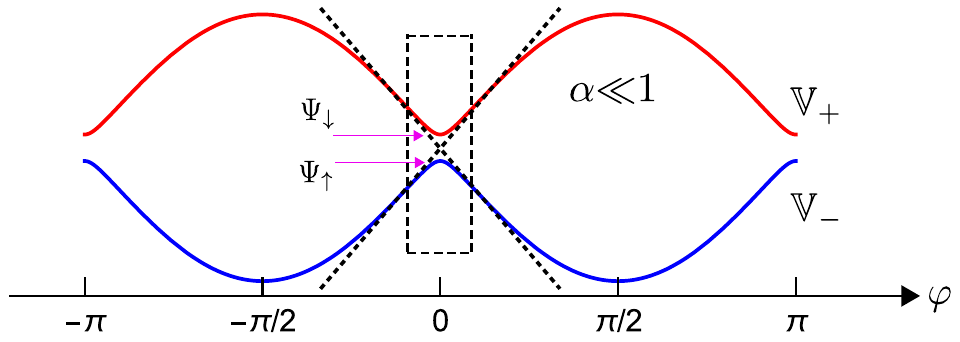}
\caption{Illustration of crossover between the perturabtive and the WKB regimes. The dashed box represents the region of linearization for $|\varphi|\ll 1$ and $h\ll v$ (see text for details).}\label{crossoverfig}
\end{center}
\end{figure}
In the previous subsection, we considered the case of sufficiently large $h$ and neglected the upper potential branch $\mathbb{V}_+$. In this case, the WKB wave function corresponding to the potential $\mathbb{V}_-$ in the limit of $h/v< |\varphi|\ll 1$ for $\varphi<0$ is given by (see appendix~\ref{wkbwfna})
\begin{align}\label{over18}
&\Psi^{\rm WKB} =\left(\frac{2^{5/4}(\sqrt2-1)}{\sqrt\pi}\right)^{1/2}\!\! \left(\frac{v}{E_C}\right)^{1/8}\left(\frac{4\left(\sqrt{2}{-}1\right)^2}{|\varphi|} \right)^{\eta}\nonumber\\
&\times\exp\Bigg[{-}2(\sqrt2{-}1)\sqrt{\frac{v}{E_C}}\Bigg]\!\exp\left[\sqrt{\frac{v}{E_C}}\left(|\varphi|{-}\frac{|\varphi|^2}{4}\right)\right].
\end{align}

For $h$ sufficiently small, the adiabatic approximation exploited in derivation of Eq.~\eqref{over18} is no longer valid. Instead, we need to solve the  Schr\"odinger equation $\mathbb{M}\Psi=E\Psi$, where the matrix $\mathbb{M}$ is defined in Eq.~\eqref{wkbeq31a} with the shift $\varphi\to\varphi-\pi/2$.  We then linearize $\mathbb{M}$ in the vicinity of $\varphi=0$ as illustrated in Fig.~\ref{crossoverfig}. For $h\ll v$, the resulting Schr\"odinger equation takes the form
\begin{equation}\label{over1}
E_C\frac{\partial^2\Psi}{\partial\varphi^2}-\left(v\varphi\sigma_z-h\sigma_x\right)\Psi=v\Psi.
\end{equation}

In the following, we will be  interested in the special limit
\begin{equation}
E_C={\rm const},\; \;h\to\infty,\;\; v\to\infty,\; \;\eta\to {\rm const}.\label{manb2}
\end{equation}
In this case, the tail of the wave function follows semiclassical approximation and the components of $\Psi$ take the similar form  of the WKB wave function~\eqref{over18}. We first focus in the linearized regime with negative $\varphi$ and define the wave function $\Psi_L$ as
\begin{equation}
\Psi_L(\varphi)=\begin{pmatrix}
\Psi_\uparrow(\varphi)\\
\Psi_\downarrow(\varphi)
\end{pmatrix},\;\;-\frac{\pi}{2}\ll\varphi< 0.\label{manbd4}
\end{equation}

Inspired by the behavior~\eqref{over18} of the wave function  at small $\varphi$, we write $\Psi_{\uparrow,\downarrow}$ as 
\begin{align}\label{over6}
\Psi_{\uparrow,\downarrow}(y)=\exp\left[-\left(v/E_C\right)^{\frac{1}{4}}y\right]\chi_{\uparrow,\downarrow}(y),
\end{align}
where we introduced the new variable $y=(v/E_C)^{1/4}\varphi$. 
We now turn to determining the form of the functions $\chi_{\uparrow,\downarrow}$. 
Using the new variables, 
Eq.~\eqref{over1} can be expressed as 
\begin{align}\label{over8}
\Psi_{\uparrow,\downarrow}(y)={-}\frac{1}{2\sqrt\eta}\left[\!\left(\!\frac{v}{E_C}\!\right)^{-\frac{1}{4}}\!\!\frac{\partial^2}{\partial y^2}\pm y{-}\left(\frac{v}{E_C}\right)^{\frac{1}{4}}\right]\Psi_{\downarrow,\uparrow}(y).
\end{align}
Substituting Eq.~(\ref{over6}), in the limit \eqref{manb2}, we obtain the following two coupled differential equations for $\chi_{\uparrow,\downarrow}$
\begin{equation}\label{over9}
\chi_{\uparrow,\downarrow}=-\frac{1}{\sqrt\eta}\left(-\frac{\partial}{\partial y}\pm\frac{y}{2}\right)\chi_{\downarrow, \uparrow}.
\end{equation}
Importantly, the right hand side of Eq.~\eqref{over9} does not contain the second derivative $\partial^2/\partial y^2$ present in Eq.~(\ref{over8}).  This significant simplification occurs under the limiting procedure (\ref{manb2}) and enables the subsequent analytic treatment of the problem. Combining the above two equations for $\chi_{\uparrow,\downarrow}$ results in the standard Weber equation
\begin{equation}\label{over10}
\frac{\partial^2\chi_\uparrow}{\partial y^2}+\left(n+\frac{1}{2}-\frac{y^2}{4}\right)\chi_\uparrow=0,\;\;\;n=-\eta.
\end{equation}

Comparing $\Psi_\uparrow$ from Eq.~\eqref{over6} and the WKB wave function $\Psi^{\rm WKB}$ given by Eq.~\eqref{over18}, we see that in the limit $|y|\gg\sqrt{\eta}$,  $\chi_\uparrow$ takes the form $\chi_\uparrow(y)\sim e^{-y^2/4}|y|^{-\eta}$. The only solution of Eq.~(\ref{over10}) that has this asymptotic behavior at $y\to -\infty$ is $D_{-\eta}(-y)$, where $D_{n}(y)$ is the standard parabolic cylinder  function. Therefore, we arrive at the result that $\chi_\uparrow(y)\propto D_{-\eta}(-y)$. Substituting this form of $\chi_\uparrow$ into the coupled differential equation~\eqref{over9} and using the properties of standard parabolic cylinder functions, we arrive at $\chi_\downarrow(y)\propto{-}\sqrt{\eta}D_{-\eta-1}(-y)$. We thus find 
\begin{equation}\label{over12}
\chi_\uparrow(y)=\mathcal{A} D_{-\eta}({-}y),\; \chi_\downarrow(y)={-}\mathcal{A}\sqrt{\eta}D_{{-}\eta{-}1}({-}y),
\end{equation}
where $\mathcal{A}$ is a constant. Substitution of $\chi_{\uparrow,\downarrow}$ from Eq.~\eqref{over12} into Eq.~\eqref{over6} finally provides the required expressions for the wave functions $\Psi_{\uparrow,\downarrow}$ in terms of yet unknown parameter $\mathcal{A}$. 

To find the coefficient $\mathcal{A}$, we compare the asymptotic behavior at $y\to -\infty$ of $\Psi_\uparrow$ given by Eqs.~\eqref{over12} and~\eqref{over6} with the WKB wave function~\eqref{over18}.  This procedure gives 
\begin{align}
\mathcal{A} &=\left(\frac{2^{\frac{5}{4}}\left(\sqrt2-1\right)}{\sqrt\pi}\right)^{\frac{1}{2}} \left(\frac{v}{E_C}\right)^{\frac{1}{8}+\frac{\eta}{4}}\times\nonumber\\
&\left[4 \left(\sqrt2-1\right)^2\right]^{\eta}\;\exp\left[-2(\sqrt2-1)\sqrt{\frac{v}{E_C}}\right].\label{over19}
\end{align}

So far we obtained the complete information of $\Psi_L$ expressed in Eq.~\eqref{manbd4}. The evaluation of the tunnel splitting in the crossover regime also requires an expression of the wave function $\Psi_R$ defined for $1\gg\varphi>0$. From the symmetry of Eq.~\eqref{over1}, it is straightforward to write $\Psi_R(\varphi)=\sigma_x\Psi_L(-\varphi)$. Having derived the expressions of $\Psi_{L/R}$, we are now in a position to calculate the tunnel splitting using the standard technique outlined in Ref.~\cite{Landau_1977}. As detailed in the appendix~\ref{tunnelsplitcross}, our final result for the tunnel splitting $\Delta^{(3)}$ in the crossover regime is given by
\begin{align}
\frac{\Delta^{(3)}}{E_C} &=8\mathcal{A}^2 \sqrt{\frac{v}{E_C}}\sqrt{\frac{\pi}{2\eta}}\frac{1}{\Gamma(\eta)},\label{over16}
\end{align}
where $\Gamma(\eta)$ is the standard Gamma function. It is readily seen that $\Delta^{(3)}$ given by Eq.~\eqref{over16} recovers the corresponding perturbative expression~\eqref{wkbeq12} in the limit of $\eta\to 0$ and also the WKB expression~\eqref{w6a} in the limit of $\eta\gg 1$. The regime $\eta\sim1$ is thus the actual crossover between the perturbative and the WKB regimes. 
\subsection{Comparison of analytical  and numerical results}
\begin{figure}[t]
\begin{center}
\includegraphics[scale=0.17]{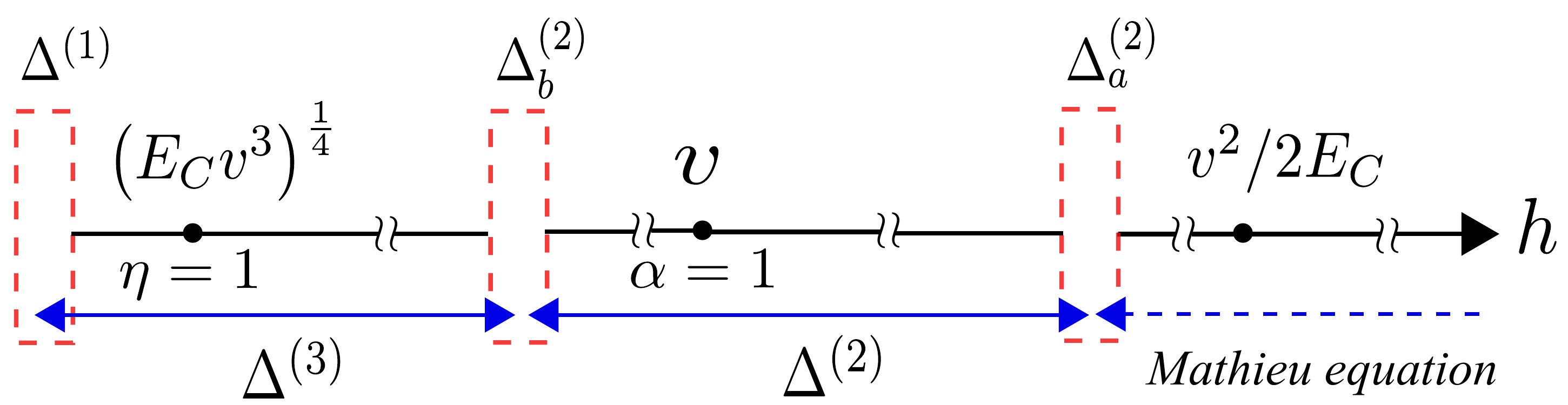}
\caption{Summary of tunnel splitting in various regimes.}\label{summaryfig}
\end{center}
\end{figure}

\begin{figure}[t]
    \centering
    \includegraphics[scale=0.35]{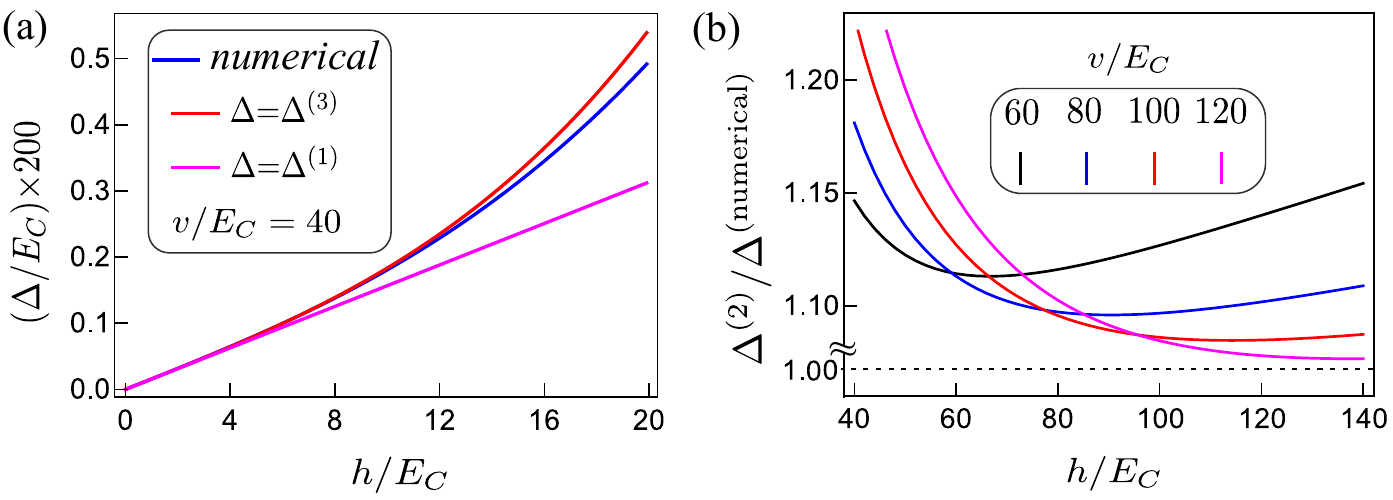}
    \caption{(a) Tunnel splitting in perturbative and crossover regimes as a function of Majorana hybridization $h$. (b) Comparison between the analytically obtained tunnel splitting in the WKB regime with that calculated numerically as a function of Majorana hybridization. Different plots show that with increasing the single particle tunneling strength $v/E_C$, WKB results asymptotically coincides with that obtained numerically.}\label{comparefig1}
\end{figure}

In the previous subsections, we analytically calculated the tunnel splitting $\Delta$, assuming $E_C \ll v$.
It encodes the dependence of ground state energy on the gate charge,
\begin{equation}\label{newEq}
E_0(n_g)=-\Delta^{(0)}\cos 2\pi n_g-\Delta\cos \pi n_g+{\rm const}.
\end{equation}
Different regimes were identified based on the relation between the strengths of Majorana overlap $h$, the single particle tunneling $v$ and the charging energy $E_C$. These three parameters are further connected by the crossover parameter $\eta$ defined by $\eta=h^2/\sqrt{16 E_C v^3}$. The perturbative regime requires $\eta\ll 1$ or, equivalently, $h\ll (E_C v^3)^{1/4}$ and corresponds to the tunnel splitting  $\Delta^{(1)}$. The WKB regime is achieved at $(E_C v^3)^{1/4} \ll h\ll v^2/E_C$. In this regime, the tunnel splitting is fully characterized by $\Delta^{(2)}$, which approaches to $\Delta^{(2)}_a$ and $\Delta^{(2)}_b$ in the limits $h\gg v$ and $h\ll v$ respectively. The perturbative regime and the WKB regime are connected by the crossover regime, in which $h\sim (E_C v^3)^{1/4}$. The tunnel splitting in the crossover regime is given by $\Delta^{(3)}$. In the case of $h\gtrsim v^2/2E_C$, our system is trivially described by the Mathieu equation. These regimes are illustrated in Fig.~\ref{summaryfig}.  
{The value of tunnel splitting $\Delta$ can be obtained numerically by evaluating the ground state energy of the Hamiltonian~(\ref{seteq0wos}) at $n_g=0$ and 1.  According to Eq.~(\ref{newEq}), $\Delta=(E_0(1)-E_0(0))/2$.  This procedure confirms our analytical results, see Fig.~\ref{comparefig1}.}

It is important to note that the splitting~\eqref{over16} has non-monotonic features as a function of the crossover parameter $\eta$. While the splitting in the vicinity of $\eta=1$ increases with $\eta$, it also manifests a decaying tail in the limit of $\eta\gg 1$. It could be difficult to numerically access the complete non-monotonic feature of tunnel splitting since the requirement of the conditions $\eta\gg 1$ and $\alpha\ll 1$ must be fulfilled simultaneously. Nevertheless, this condition might actually be relevant in the context of real experimental setups.

\section{Emergent Josephson effect}\label{blockade}

In Sec.~\ref{sec:an} we focused on the analytical calculation of the lowest energy level at $E_J = 0$ in the limit of large $v$ as a function of MZMs hybridization $h$. The general case of finite $(h, v, E_J)$ can be easily studied numerically following the discussion of the section~\ref{dpsbc}, and certain features of the spectrum have been already discussed in Sec.~\ref{qfeatures}. In the following, we focus on an interesting situation that arises when $h$ is the largest energy scale and all parameters $(h, v, E_J)$ are finite.

\begin{figure}[t]
   \begin{center}
    \includegraphics[scale=0.47]{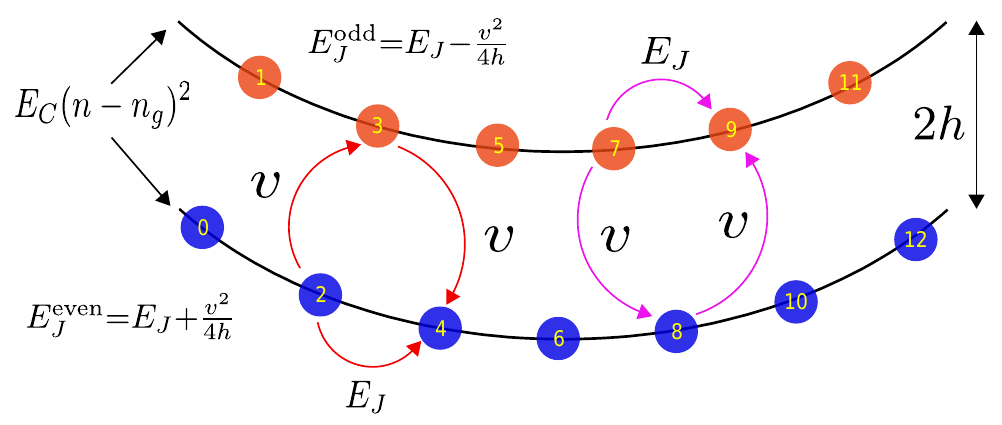}
        \caption{Illustration of energy parabolas in odd and even sectors. When Majorana hybridization becomes the largest energy scale, virtual tunneling between the two sectors produces Josephson-like coupling. The sign of this coupling is opposite in the two sectors.}\label{emergentJJ}
    \end{center}
\end{figure}

As discussed previously, the two sectors that correspond to the even and odd $n$ on the island get effectively decoupled when Majorana hybridization $h$ exceeds all other energy scales. In this case, the Majorana tunneling events amount to virtual processes as illustrated in Fig.~\ref{emergentJJ}. These processes induce a contribution to effective Josephson coupling in each sector. As mentioned in the subsection~\ref{wkbregm}, when $h \gg v$, the strength of effective induced Josephson coupling is $\sim v^2/h$. Interestingly, the sign of this contribution is the opposite in even and odd $n$ sectors since the intermediate state has a higher energy than the initial state in the even sector and lower in the odd. Therefore, the induced Josephson coupling is given by
\begin{equation}
E_J^{\rm even, odd}=E_J\pm\frac{v^2}{4h}.\label{manbd5}
\end{equation}

Because the two sectors are essentially decoupled in the large $h$ limit, one can  define the pseudo ground state in the odd $n$ sector, in addition to the true ground state in the  even parity sector. In the pseudo-ground state, the induced Josephson coupling can offset, nullify, or switch the sign of the Josephson contribution $E_J$ originating from the tunneling of the Cooper pairs. The sign change of $E_J$ makes the standard Josephson junction into a $\pi$- junction~\cite{Ioffe1999,Ioffe_2001}, a desirable element of some superconducting quantum circuits that naturally leads to bistability.

This effect could also  be useful for  sensing applications, including single microwave photon detection. This can be achieved by coupling a microwave photon field to the MZM tunneling $v$ via gate, as in the gatemon qubit~\cite{gatemon}. From Eq.~\eqref{wkbeq31a} it is seen that such modulation will enable photon-assisted transitions between the ground and the pseudo-ground states of the system, when the photon energy matches their splitting, $2h$. Transition into pseudo-ground state reduces the Josephson critical current on the junction; thus a possible detection scheme  corresponds to current-biasing the junction below the value of the critical current in the ground-state but above the one in the pseudo-ground state. Then, photon absorption in the junction would cause the junction to jump into the resisting state, where it would stay until the system is reset back into the ground state.

We finally note that the emergent Josephson physics expressed by Eq.~\eqref{manbd5} arises due to the blockade of MZMs tunnelling because of large parity-splitting energy. This effect is very particular to MZMs, and  therefore can be a  useful indicator for the presence of MZMs in real experimental systems, complementary to other proposed methods as in Ref.~\cite{Prada_Aguado_2020b}. We note that related parity-sensitive effects have been discussed recently in a rather different setup hosting Majorana Kramers pairs~\cite{Fu_2018, Fu_2022}.

\section{Conclusion}\label{dissoutlook}
To conclude, we developed an analytical as well as numerical method for solving energy spectra of the Majorana-SC qubit hybrid. We showed that various competing effects in this hybrid device result in several operational regimes. Unique features associated with different regimes have been explored using both qualitative and quantitative approaches. Moreover, we studied the crossover among these regimes and derived the compact expressions for the crossover scales. 

We demonstrated that the competition between intra-wire ($h$) and inter-wire ($v$) Majorana tunneling gives rise to the ``Majorana-blockade" effect. This effect may serve as a smoking-gun for the presence of Majorana modes and can also be exploited for sensing applications.

The main qualitative feature of the Majorana-SC qubit hybrid is the possibility of the coherent single electron tunneling between the island and the lead. 
It allows the system's energy be a periodic function of the gate charge $n_g$ with period 1 when $E_J = h = 0$ (as compared to the period 2 for the conventional superconducting devices). This feature parallels the doubling of the period in the current-phase relationship, which is considered one of the hallmarks of Josephson junctions containing MZMs~\cite{Beenakker_2013}. It has been recognized however that the observation of this effect requires the junction parity to be conserved on the timescale of the experiment. 
Indeed, any hybridization of the junction MZMs ($\gamma_2$ and $\gamma_3$ in Fig. \ref{setup1}) with the ``external" MZMs ($\gamma_1$ and $\gamma_4$)  
allows the junction parity $i\gamma_2\gamma_3$ to adjust to minimize energy, resulting in the standard periodicity of the current-phase relationship.   
In contrast, from Eq.~\eqref{energygnd}, we find that the single-charge (Majorana) periodicity dominates up to a finite $h$, given by Eq. (\ref{hc}). The fact that a sufficiently large $h$ is necessary to wash out the Majorana signatures, highlights the importance of the finite charging energy, which introduces dynamics. {Such dynamics is absent in the derivation of the current-phase relationship in which case the charging energy is set to zero}.

\section*{Acknowledgements}
We are thankful to A. Shnirman, M. Vavilov, and J. Koch for fruitful discussions.
This work was supported by the US Department of Energy, Office of Science, Basic Energy Sciences, Materials Sciences and Engineering Division.

\appendix

\section{Equivalence between the approach of Sec.~\ref{sec:phH} and Ref.~\cite{Ginossar_2014}}
\label{app:33}
In this appendix we show the relationship between our approach presented in the section~\ref{sec:phH} and the approach of Ref.~\cite{Ginossar_2014}. 
To this end, we chose the transformation between  the electron number and phase space in the form 
\begin{equation}
\psi_n=\frac{1}{\sqrt{2\pi}}\int^\pi_{-\pi}d\varphi\psi(\varphi) e^{-in\varphi}.\label{wkbeq20}
\end{equation}
Unlike the main text, here we do not include the $n_g$ shift in the exponent of Eq.~\eqref{wkbeq20}, and thus the standard boundary condition applies, i.e., $\psi(\varphi+2\pi)=\psi(\varphi)$. In terms of the function $\psi(\phi)$, the Schr\"odinger equation~\eqref{eq:SEn} can be written as
\begin{align}
E \psi(\varphi) =&\left[E_C({-i\partial_\varphi} - n_g)^2-v\cos\varphi-E_J\cos 2 \varphi\right]
\psi(\varphi)\nonumber\\
&-h\psi\left(\varphi-\pi{\rm sgn}\varphi\right).\label{eq:SEphi2}
\end{align}
%{where we assume $-\pi<\varphi<\pi$.}

When the gate voltage is explicitly included into the Hamiltonian, two components of the function $\Psi$ defined in Eq.~\eqref{twocompf} satisfy the boundary conditions
\begin{align}\label{eq:bc2}
\Psi_\uparrow(\varphi+\pi)&=\Psi_\downarrow(\varphi) ,\nonumber\\
\Psi_\downarrow(\varphi+\pi)&=\Psi_\uparrow(\varphi) .
\end{align}
Equation.~\eqref{eq:SEphi2} then takes the form
\begin{align}
\mathbb{L}\Psi=E\Psi,\label{wkbeq37}
\end{align}
where 
\begin{equation}
\mathbb{L}=E_C({-i\partial_\varphi} - n_g)^2-E_J\cos 2\varphi+v\cos\varphi\sigma_z-h\sigma_x.\label{wkbeq31d}
\end{equation}

The combination of Hamiltonian~\eqref{wkbeq31d} and boundary conditions (\ref{eq:bc2}) is physically equivalent to the combination of Hamiltonian and boundary conditions in Sec.~\ref{sec:phH}. However, the former is more convenient for demonstrating the correspondence with the approach  presented in Ref.~\cite{Ginossar_2014}. 

Indeed, we can define the unitary matrix
\begin{eqnarray}
    \mathbb{U}=\frac{1}{\sqrt2}\begin{pmatrix}
        1 & \phantom{-}1\\
        1 &-1
    \end{pmatrix},
\end{eqnarray}
which rotates the basis into
\begin{equation}
    \begin{pmatrix}
        \Psi_{\rm e}\\
        \Psi_{\rm o}
    \end{pmatrix}=\mathbb{U}\begin{pmatrix}
        \Psi_\uparrow\\
        \Psi_\downarrow
    \end{pmatrix}.
\end{equation}
The rotated Hamiltonian $\mathbb{L}'=\mathbb{U}\mathbb{L} \mathbb{U}^\dagger$  is equivalent to the Hamiltonian studied in Ref.~\cite{Ginossar_2014} after the trivial modification $\varphi\to 2\varphi$. This difference accounts for the fact that in this work we measure charge in the units of single electrons, while Ref.~\cite{Ginossar_2014} counts charge in units of Cooper pairs; this also results in  $n_g\to n_g/2$. The basis $\Psi_{\rm e/o}$ satisfies respectively the periodic and anti-periodic boundary conditions with period $\pi$, i.e., $\Psi_{\rm e/o}(\varphi+\pi)=\pm \Psi_{\rm e/o}(\varphi)$. This establishes a formal equivalence of our treatment and the one of the Ref.~\cite{Ginossar_2014}.

\section{Derivation of tunnel splitting in perturbative regime}\label{app1}
In this appendix, we outline the derivation of Eq.~\eqref{wkbeq12} starting from Eq.~\eqref{wkbeq5aa}. For the sake of simplicity of presentation, we introduce the wave functions $\Psi_{1,2}$ defined by $\psi_0(x)=\Psi_1(x)$ and $\psi_0(x-\pi)=\Psi_2(x)$. The potentials corresponding to $\Psi_{1, 2}$ then take the form
\begin{equation}
\mathbb{V}_{1, 2}(x)=v\left(1\pm\cos x\right),\label{ap2}
\end{equation}
where we added the constant $v$ to $\mathbb{V}_{1, 2}$. The wave functions $\Psi_{1, 2}$ under the semiclassical WKB approximation can be written as
\begin{align}
\Psi_1(x)&=\left(\frac{\Omega^2}{4\pi e}\right)^{1/4}\!\!\left(\!\frac{m}{|p_{1}(x)|}\!\right)^{1/2}
\!\!\exp\Bigg[\!{-}\frac{1}{\hbar}\int^{a_1}_x \!\!dy\;|p_{1}(y)|\Bigg],\nonumber\\
\Psi_2(x)&=\left(\frac{\Omega^2}{4\pi e}\right)^{1/4}\!\!\left(\!\frac{m}{|p_{2}(x)|}\!\right)^{1/2}
\!\!\exp\Bigg[\!{-}\frac{1}{\hbar}\int_{a_2}^x \!\!dy\;|p_{2}(y)|\Bigg].\label{ap1}
\end{align}
To arrive at Eq.~\eqref{ap1} we exploited the fact that near the minima of potentials~\eqref{ap2}, the corresponding wave functions are well approximated by the ground state harmonic oscillator wave functions~\cite{Landau_1977}. The frequency $\Omega$ of small amplitude oscillation around the minima of potentials~\eqref{ap2} is given by $m\Omega^2=v$, where $m=\hbar^2/2E_C$. The classical turning points $a_{1, 2}$ are defined in terms of the parameter $z$ satisfying $z^2\simeq \hbar/m\Omega$ as
\begin{equation}
a_1=\pi-z, \;\;\;a_2=z,\;\;z\ll 1.
\end{equation}
The semiclassical momenta $p_\sigma$ in Eq.~\eqref{ap1} can be expressed in the form
 \begin{equation}
p_{\sigma}(x)=\sqrt{2m\left[\mathbb{V}_\sigma(x)-\mathbb{V}_\sigma(x=a_\sigma)\right]},\;\;\sigma=1, 2.\label{ap3}
\end{equation}

In the limit of $v\gg E_C$, the leading order contribution to the tunnel splitting given by Eq.~\eqref{wkbeq5a} takes the form
\begin{align}
\Delta^{(1)}=2mh\left(\frac{\Omega^2}{4\pi e}\right)^{1/2} \!\!\!\int^\pi
_0\!\!\! \frac{dx}{\sqrt{|p_{1}(x)||p_{2}(x)|}}\exp\left[{-}\frac{\mathcal{F}(x)}{\hbar}\right],\label{ap4}
\end{align}
where the function $\mathcal{F}$ is defined by
\begin{equation}\label{ap5}
\mathcal{F}(x)=-\int_{a_1}^x dy\;|p_{1}(y)|+\int_{a_2}^x dy\;|p_{2}(y)|.
\end{equation}
To evaluate the integral in Eq.~\eqref{ap5}, we proceed with the saddle point method. Assuming that the integrand of Eq.~\eqref{ap5} becomes maximum at $x=\pi/2$, we write the saddle point solution for $\mathcal{F}$, which takes the form
\begin{align}\label{ap6}
\mathcal{F}(x)&\simeq\mathcal{F}\left(\frac{\pi}{2}\right)+\frac{\hbar}{ z^2\sqrt2}\left(x-\frac{\pi}{2}\right)^2.
\end{align}
From equations~\eqref{ap4} and~\eqref{ap6}, we obtain the expression for $\Delta^{(1)}$ to the leading order in $z$ 
\begin{align}
\Delta^{(1)}
&=h\frac{2^{1/4}z}{\sqrt{2e}}\exp\left[-\frac{\mathcal{F}(\pi/2)}{\hbar}\right].\label{ap7}
\end{align}
In the limit of $z\ll 1$, the evaluation of the exponential factor in the above equation yields
\begin{align}\label{ap8}
\exp\left[\!{-}\frac{\mathcal{F}(\frac{\pi}{2})}{\hbar}\!\right]\!=\!\frac{8\sqrt{e}}{z(\sqrt2{+}1)}\exp\!\left[\!{-}4(\sqrt2{-}1)\sqrt{\frac{v}{E_C}}\right].
\end{align}
Using equations~\eqref{ap8} and Eq.~\eqref{ap7}, we obtain the required expression of tunnel splitting in the perturbative regime, which is quoted in the main text Eq.~\eqref{wkbeq12}.

\section{Derivation of tunnel splitting in WKB regime}\label{wregime}
For the evaluation of tunnel splitting formula in WKB regime, we focus in the double-well regime, as shown in Fig.~\ref{wkbfig}, of the potential $\mathbb{V}_-$ defined in Eq.~\eqref{w1}. The frequency $\omega$ of small amplitude oscillation around the minimum $\varphi=\pi/2$ of the potential $\mathbb{V}_-$ is given by
\begin{align}
\omega^2=\left.\frac{d^2\mathbb{V}_-(\varphi)}{d\varphi^2}\right|_{\varphi=\pi/2}=\frac{v}{m}\frac{1}{\sqrt{1{+}\alpha^2}},\;\;m=\frac{\hbar^2}{2E_C}.\label{ppd}
\end{align}
The potential $\mathbb{V}_-$ has left and right classical turning points at $\varphi=\pm\pi/2\mp l$, where potential coincides with the ground state energy;
\begin{equation}
\mathbb{V}_-(\varphi=\pm\pi/2\mp l)=E_0=\frac{1}{2}\hbar\omega.
\end{equation}
For $l\ll \pi/2$, we use the semi-classical approximation $l^2\simeq \hbar/m\omega$. In this case, the tunnel splitting corresponding to the potential $\mathbb{V}_-$ in Eq.~\eqref{w1} is given by~\cite{Landau_1977,Garg_2000}
\begin{align}
\frac{\Delta^{(2)}}{E_C}=\frac{\hbar\omega}{E_C\sqrt{e\pi}}\exp\left[-\sqrt{\frac{v}{E_C}}\int^{\frac{\pi}{2}-l}_{-\frac{\pi}{2}+l}d\varphi\; U(\varphi)\right],\label{w2}
\end{align}
where
\begin{equation}
U(\varphi)=\sqrt{\sqrt{\alpha^2+(\cos l)^2}-\sqrt{\alpha^2+(\sin\varphi)^2}}.
\end{equation}
In the strong barrier limit $l\ll\pi/2$, exploiting the fact that $\mathbb{V}_-(\varphi=\pm\pi/2)=0$, the Eq.~\eqref{w2} can be expressed into the form~\cite{Garg_2000}
\begin{align}
&\frac{\Delta^{(2)} }{E_C}=\frac{\hbar\omega}{E_C}\left(\frac{m\omega\pi}{\hbar}\right)^{1/2}\exp\mathcal{S}_0\;\exp\mathcal{S}_1,\label{degrg1}
\end{align}
where
\begin{align}
\mathcal{S}_0& =-\frac{1}{\hbar}\int^{\pi/2}_{-\pi/2}\sqrt{2m\mathbb{V}_-(\varphi)}\;d\varphi,\\
\mathcal{S}_1&=\lim_{\delta\to 0^+}\int^{\pi/2-\delta}_0\left(\frac{m\omega}{\sqrt{2m\mathbb{V}_-(\varphi)}}{-}\frac{1}{\frac{\pi}{2}{-}\varphi}\right)\;d\varphi.\label{degrg3}
\end{align}
We calculate these integrals and substitute them into Eq.~\eqref{degrg1} to arrive at the expression of tunnel splitting~\eqref{w3}.

\section{Derivation of the WKB wave function}\label{wkbwfna}
The WKB wave function corresponding to the potential $\mathbb{V}_-$, which is sketched in Fig.~\ref{wkbfig}, for $\varphi\leq 0$  can be written as~\cite{Landau_1977}
\begin{equation}
\Psi^{\rm WKB}(\varphi)=\left(\frac{\omega^2}{4\pi e}\right)^{1/4}\left(\frac{m}{|p(\varphi)|}\right)^{1/2}\exp\mathscr{M}(\varphi),\label{aw1}
\end{equation}
where the function $\mathscr{M}$ is defined by
\begin{equation}
\mathscr{M}(\varphi)= -\frac{1}{\hbar}\int^{\varphi}_{-\frac{\pi}{2}+l}\;dx |p(x)|.\label{aw2}
\end{equation}
Here the semiclassical momentum $p(x)$ takes the usual form
\begin{equation}
p(x)=\sqrt{2m\left(\mathbb{V}_-(x)-\mathbb{V}_-(x=-\frac{\pi}{2}+l)\right)}.
\end{equation}
We now express $\mathscr{M}$ into the form
\begin{equation}
\mathscr{M}=\mathscr{M}_0+\mathscr{M}_1,\label{aw6}
\end{equation}
with
\begin{align}\mathscr{M}_0(\varphi) &=-\sqrt{\frac{v}{E_C}}\int^{\varphi}_{-\frac{\pi}{2}+l}\;dx\sqrt{\cos l-|\sin x|},\\
\mathscr{M}_1(\varphi)&=-\sqrt{\frac{v}{E_C}}\int^{\varphi}_{-\frac{\pi}{2}+l}\;dx\Bigg[-\sqrt{\cos l-|\sin x|}\nonumber\\
&+\sqrt{\sqrt{\alpha^2+(\cos l)^2}-\sqrt{\alpha^2+(\sin x)^2}}\Bigg].
\end{align}
To the leading order in the small parameters $\alpha$ and $l$, $\mathscr{M}_0$ and $\mathscr{M}_1$ in the limit of $\alpha<|\varphi|\ll 1$ take the form
\begin{align}
\mathscr{M}_0(\varphi)&=-\sqrt{\frac{v}{E_C}}\Bigg[2(\sqrt2-1)-|\varphi|+\frac{|\varphi|^2}{4}\nonumber\\
&\;\;\;\;\;\;-\frac{l^2}{4\sqrt2}-\frac{l^2}{2\sqrt2}\log\left(\frac{8(\sqrt2-1)}{l}\right)\Bigg],\label{aw3}
\end{align}
and
\begin{equation}
\mathscr{M}_1(\varphi)= \eta \log \left(\frac{4 \left(\sqrt{2}-1\right)^2}{|\varphi|}\right).\label{aw4}
\end{equation}
For $\alpha\ll 1$ and $l\ll 1$, we make further approximations
\begin{align}
&\frac{l^2}{\sqrt2}\sqrt{\frac{v}{E_C}} = 1,\;\omega^2=\sqrt{\frac{2E_C v}{\hbar^2}},\;\frac{m}{|p(\varphi)|}= \frac{\hbar}{2}\sqrt{\frac{1}{E_C v}}.\label{aw5}
\end{align}
Substituting Eqs.~\eqref{aw6} and~\eqref{aw3}$-$\eqref{aw5} into Eq.~\eqref{aw1}, we obtain the required expression for the wave function quoted in the main text Eq.~\eqref{over18}.

\section{Tunnel splitting in the crossover regime}\label{tunnelsplitcross}
To apply the standard method of evaluating tunnel splitting outlined in Ref.~\cite{Landau_1977} to our problem, we first need to generalize it for the case of two-component wave functions $\Psi_{L/R}$. To this end, we define the symmetric and anti-symmetric combinations of the wave functions $\Psi_{L/R}$ as
\begin{equation}
\Psi_{\pm}(\varphi)=\frac{\Psi_L(\varphi)\pm\Psi_R(\varphi)}{\sqrt2},
\end{equation}
with corresponding energies $\epsilon_{\pm}$. Since both $\Psi_L$ and $\Psi_+$ follow the Eq.~\eqref{over1}, we write \begin{equation}
E_C\frac{\partial^2\Psi_L}{\partial \varphi^2}-\left(v\varphi\sigma_z-h\sigma_x\right)\Psi_L=v\Psi_L,\label{ts1}
\end{equation}
and
\begin{equation}
E_C\frac{\partial^2\Psi_+}{\partial \varphi^2}-\left(v\varphi\sigma_z-h\sigma_x\right)\Psi_+=\epsilon_+\Psi_+.\label{ts2}
\end{equation}
We proceed further by multiplying Eq.~\eqref{ts1} from left by $\Psi_+$, Eq.~\eqref{ts2} from left by $\Psi_L$ and subtracting the resulting expressions. Thus obtained result upon integrating from $-\infty$ to $0$ gives
\begin{align}\label{ts3}
\frac{v-\epsilon_+}{E_C}&=\sqrt2\int^0_{-\infty}\!\!\!\!d\varphi\left(\Psi_+\cdot\frac{\partial^2\Psi_L}{\partial \varphi^2}{-}\Psi_L\cdot\frac{\partial^2\Psi_+}{\partial \varphi^2}\right)\nonumber\\
&=2\left.\left(\Psi_\downarrow\frac{\partial\Psi_\uparrow}{\partial \varphi}+\Psi_\uparrow\frac{\partial\Psi_\downarrow}{\partial \varphi}\right)\right|_{\varphi=0}.
\end{align}
To arrive at Eq.~\eqref{ts3}, we exploited the identity
\begin{equation}
\int^0_{-\infty} d\varphi\Psi_L.\Psi_+\simeq\frac{1}{\sqrt2}.
\end{equation}
The evaluation of $(v-\epsilon_-)/E_C$ proceeds similarly and thus the tunnel splitting $\Delta^{(3)}\equiv\epsilon_--\epsilon_+$ can be written as
\begin{equation}
\frac{\Delta^{(3)}}{E_C}=4\left.\left(\Psi_\downarrow\frac{\partial\Psi_\uparrow}{\partial \varphi}+\Psi_\uparrow\frac{\partial\Psi_\downarrow}{\partial \varphi}\right)\right|_{\varphi=0}.\label{ma3}
\end{equation}
From Eqs.~\eqref{over6} and~\eqref{over12}, we have for the expression of $\Psi_\uparrow$ in the form
\begin{align}
&\Psi_\uparrow(\varphi)=\mathcal{A}\exp\left({-}\sqrt{\frac{v}{E_C}}\varphi\right)D_{-\eta}\left[\!{-}\left(\frac{v}{E_C}\right)^{\frac{1}{4}}\!\!\varphi\right].\label{ma1}
\end{align}
Similarly, $\Psi_\downarrow$ can be written as
\begin{align}
\Psi_\downarrow(\varphi)=&-\mathcal{A}\exp\left(-\sqrt{\frac{v}{E_C}}\varphi\right)\sqrt\eta \nonumber\\
&\times D_{-\eta-1}\left[-\left(\frac{v}{E_C}\right)^{\frac{1}{4}}\varphi\right].\label{ma2}
\end{align}
Substitution of Eqs.~\eqref{ma1} and~\eqref{ma2} into Eq.~\eqref{ma3} gives the required formula for tunnel splitting in the crossover regime. In the regime of our interest $v\gg E_C$, the main contribution to the derivatives in Eq.~\eqref{ma3} comes from the exponential factor of the wave functions $\Psi_{\uparrow,\downarrow}$. Therefore, the leading order term of Eq.~\eqref{ma3} can be evaluated by neglecting the derivative of parabolic cylinder functions. The resulting expression of tunnel splitting takes the compact form given by
\begin{align}
\frac{\Delta^{(3)}}{E_C} &=8\mathcal{A}^2\sqrt{\frac{v}{E_C}} D_{-\eta}(0) \sqrt\eta D_{-\eta-1}(0),\label{ma4}
\end{align}
Using the properties of parabolic cylinder functions, Eq.~\eqref{ma4} can equivalently be expressed in the form of Eq.~\eqref{over16}.
%\bibliography{biblio.bib}
\end{document}